\documentclass[preprint,superscriptaddress,preprintnumbers,amsmath,amssymb,pra]{revtex4}
\usepackage{graphicx}

\begin{document}
\def\Xint#1{\mathchoice
   {\XXint\displaystyle\textstyle{#1}}%
   {\XXint\textstyle\scriptstyle{#1}}%
   {\XXint\scriptstyle\scriptscriptstyle{#1}}%
   {\XXint\scriptscriptstyle\scriptscriptstyle{#1}}%
   \!\int}
\def\XXint#1#2#3{{\setbox0=\hbox{$#1{#2#3}{\int}$}
     \vcenter{\hbox{$#2#3$}}\kern-.5\wd0}}
\def\ddashint{\Xint=}
\def\dashint{\Xint-}

\newcommand{\sil}{\sigma_{\|}}
\newcommand{\sit}{\sigma_{\bot}}
\newcommand{\sila}{\sigma_{\|}(\omega,k)}
\newcommand{\sita}{\sigma_{\bot}(\omega,k)}
\newcommand{\sig}{\sigma(\omega)}
\newcommand{\ar}{(\omega,k)}
\newcommand{\ho}{\hbar\omega}

\thispagestyle{empty}

\title{Impact of chemical potential on the reflectance of graphene in the
infrared and microwave domains}

\author{
G.~L.~Klimchitskaya}
\affiliation{Central Astronomical Observatory at Pulkovo of the
Russian Academy of Sciences, Saint Petersburg,
196140, Russia}
\affiliation{Institute of Physics, Nanotechnology and
Telecommunications, Peter the Great Saint Petersburg
Polytechnic University, Saint Petersburg, 195251, Russia}

\author{
V.~M.~Mostepanenko}
\affiliation{Central Astronomical Observatory at Pulkovo of the
Russian Academy of Sciences, Saint Petersburg,
196140, Russia}
\affiliation{Institute of Physics, Nanotechnology and
Telecommunications, Peter the Great Saint Petersburg
Polytechnic University, Saint Petersburg, 195251, Russia}
\affiliation{Kazan Federal University, Kazan, 420008, Russia}

\author{
V.~M.~Petrov}
\affiliation{Institute of Advanced Manufacturing Technologies,
Peter the Great Saint Petersburg
Polytechnic University, Saint Petersburg, 195251, Russia}

\begin{abstract}
   The reflectance of graphene is investigated in the framework
of the Dirac model with account of its realistic properties, such
as nonzero chemical potential and band gap, at any temperature.
For this purpose, the exact reflection coefficients of the
electromagnetic waves on a graphene sheet expressed via the
polarization tensor and ultimately via the electrical conductivity
of graphene have been used. The reflectance of graphene is computed
as a function of frequency and chemical potential at different
temperatures and values of the band-gap parameter. The minimum
values of the reflectance are found which are reached in the
infrared domain at the points of vanishing imaginary part of the
conductivity of graphene. For a gapped graphene, the maximum
reflectance equal to unity is reached at the points where the
imaginary part of conductivity diverges. The computational
results demonstrate an interesting interplay between the band
gap and chemical potential in their combined effect on the
reflectance. Specifically, there are wide frequency intervals
where the reflectance of graphene increases with increasing
chemical potential and decreasing band gap. The numerical
computations are found to be in good agreement with the analytic
asymptotic expressions in the regions of their applicability.
Several technological areas, where the obtained results could be
used, are listed.
\end{abstract}

\maketitle

\section{INTRODUCTION}

   Currently the reflectivity properties of graphene have
been investigated both theoretically and experimentally using
a variety of approaches and techniques \cite{1,2}. In the
majority of cases, studies on the optics of graphene were
based on the investigation of its electrical conductivity.
On this subject a great number of results have been obtained
using the Kubo formalism, the Boltzmann transport theory, the
current-current correlation functions, the two-dimensional
Drude model, and others (see the review papers \cite{3,4,5}).
Although some of these approaches employ simple intuitive
models and are of phenomenological character, they were
fruitfully used to investigate the reflectivity of graphene
in different spectral domains \cite{6,7,8,9,10,11,12,13}.
In addition to the academic interest, a knowledge of the
reflectivity properties is required in prospective
technological applications of graphene, such as in the
modulators, detectors and switches \cite{11,13,14}, solar
cells \cite{15}, corrosion protection \cite{16}, transparent
electrodes \cite{17} etc.

A more complete theory of the optical properties of graphene
at relatively low frequencies in the microwave and infrared
domains would be beneficial for both fundamental physics
and its applications. Fortunately, at these frequencies
(i.e., at energies below 1--2~eV) graphene is well described
by the Dirac model. This model assumes that graphene
quasiparticles are massless or very light and obey the
(2+1)-dimensional Dirac equation, where the Fermi velocity
$v_F = c$/300 plays the role of the speed of light $c$
\cite{1,2,18}. In the framework of the Dirac model, it is
possible to obtain an exact expression for the polarization
tensor of graphene on the basis of first principles of
quantum electrodynamics at nonzero temperature. This tensor
describes a response of graphene to the electromagnetic
field and, thus, can be used for a unified theoretical
description of the broad range of physical phenomena
including the Casimir and van der Waals forces, electrical
conductivity and optical properties. Taking into account
that the Dirac model incorporates naturally a nonzero
band gap $\Delta$ in the energy spectrum of quasiparticles
and a chemical potential $\mu$, the polarization tensor
provides the reliable foundation for a description of
physical processes in real (doped) graphene samples at
energies below 1--2~eV.
At higher energies, i.e., beyond the Dirac model, the
conductivity and reflectivity properties of the graphene-like
materials have been studied with account of Coulomb interaction,
interactions with phonons, and a magnetic field
\cite{2,7,18, 18a,18b,18c,18d,18e,18f}.

The exact expression for the polarization tensor of
graphene at zero temperature in the one-loop approximation
was derived in Ref.~\cite{19} (see also Ref. \cite{20} and
literature therein for some special cases considered
previously). In Ref.~\cite{21} the polarization tensor at
nonzero temperature has been found, but only at the
pure imaginary Matsubara frequencies. The polarization tensor
of Ref.~\cite{21} was used to derive the Lifshitz-type formula,
where the reflection coefficients are expressed via this tensor,
and to investigate the fluctuation-induced van der Waals and
Casimir forces in graphene systems
\cite{21,22,23,24,25,26,27,28,29,30,31,32}.
It was not possible, however, to calculate the electrical
conductivity and optical properties of graphene which are expressed
via the polarization tensor defined at real frequences.
Another representation for
the polarization tensor of graphene at nonzero temperature,
allowing an analytic continuation to the entire complex
frequency plane, was found in Ref.~\cite{33}. It is significant
that at the pure imaginary Matsubara frequencies the
representations of Refs.~\cite{21} and \cite{33} coincide,
but take dissimilar values at all other frequencies. A more
universal representation of Ref. \cite{33} was used in
calculations of the Casimir force \cite{34,35,36,37} and,
after a continuation to the axis of real frequencies, for
investigation of the electrical conductivity \cite{38,39}
and reflectances \cite{40,41,42} of an undoped graphene.

The polarization tensor of doped graphene with a nonzero
chemical potential has been derived in Ref.~\cite{43} using
the representation of Ref.~\cite{33}. This tensor was used
to study an impact of nonzero chemical potential on the
thermal Casimir force in graphene systems Refs.~\cite{44,45}
and, after a continuation to the axis of real frequencies,
on the electrical conductivity of graphene \cite{46}. As a
result, the theory of electrical conductivity of graphene
has been developed in the framework of the Dirac model,
which takes into a complete account the effect of the nonzero
band gap and chemical potential. The derived expressions for
 real and imaginary parts of the conductivity of graphene
were found to be in agreement with the causality conditions
and satisfy the Kramers-Kronig relations \cite{47}. This
solved the problem which was under a discussion in the
previous literature \cite{48,49,50,51}.

In this paper, the recently developed theory is applied to
investigate an impact of chemical potential on the
reflectance of gapped graphene in the application region of
the Dirac model, i.e., in the infrared and microwave domains.
We present the exact reflection coefficients and reflectance
for the transverse magnetic and transverse electric (i.e.,
$p$- and $s$-polarized) electromagnetic waves on
a graphene sheet expressed via the polarization tensor or
via the conductivity. An explicit expression for the angular
dependence of graphene reflectance, giving the major
contribution to the results, is also obtained. The reflectance
of graphene at the normal incidence is investigated as a
function of frequency and chemical potential at different
temperatures $T$ and values of the band-gap parameter. It is shown
that the minimum reflectance is reached at the values of
frequency and chemical potential where the imaginary part of
the conductivity of  graphene vanishes. In doing so, the minimum
reflectance either vanishes or takes rather small value
depending on the magnitude of the real part of conductivity.
At different temperatures a qualitatively different impact of
the chemical potential on the reflectance of graphene is
observed. The computational results demonstrate an interesting
interplay between the band gap and chemical potential in their
combined effect on the reflectance. There are rather wide
frequency intervals where the reflectance increases with
increasing chemical potential and decreasing band gap.

The paper is organized as follows. In Sec.~II, we present
general expressions for the reflectance of graphene in the
framework of the Dirac model in the case of nonzero band gap
and chemical potential. In Sec.~III, the impact of chemical
potential on the reflectance is investigated for a gapless
graphene. Section IV is devoted to an interplay in the impact
of nonzero band gap and chemical potential of graphene on its
reflectance. In Sec.~V, the reader will find our conclusions
and discussion.

\section{Reflectance of graphene in the framework of the Dirac model}

   We consider a plane wave of frequency $\omega$ incident on
the graphene sheet under the angle $\theta_i$. For real photons
on a mass shell the magnitude of the wave-vector component
parallel to graphene is $k = \omega \sin\theta_i /c$. Then the
amplitude reflection coefficients for the transverse magnetic
(TM) and transverse electric (TE) polarizations of the
electromagnetic field along the real frequency axis are
given by \cite{33,41}
\begin{eqnarray}
&&
r_{\rm TM}(\omega,\theta_i)=
\frac{\omega\cos\theta_i\Pi_{00}(\omega,k)}{2i\hbar ck^2+
\omega\cos\theta_i\Pi_{00}(\omega,k)},
\nonumber \\
&&
r_{\rm TE}(\omega,\theta_i)=
\frac{c\Pi(\omega,k)}{2i\hbar \omega k^2\cos\theta_i-
c\Pi(\omega,k)}.
\label{eq1}
\end{eqnarray}
\noindent
Here, $\Pi_{00}$ is the 00 component of the polarization tensor of graphene
$\Pi_{\mu\nu}$, $\mu,\,\nu=0,\,1,\,2$ and
\begin{equation}
\Pi(\omega,k)=k^2\Pi_{\rm tr}(\omega,k)+
\left(\frac{\omega^2}{c^2}-k^2\right)\Pi_{00}(\omega,k),
\label{eq2}
\end{equation}
\noindent
where $\Pi_{\rm tr}\equiv\Pi_{\mu}^{\,\mu}$ is the trace of this tensor.
Note that in calculations of the Casimir force between two graphene sheets
one should replace $\omega$ in Eq.~(\ref{eq1}) with the pure imaginary Matsubara
frequencies
$i\xi_l=2\pi ik_BTl/\hbar$, where $k_B$ is the Boltzmann constant
and $l=0,\,1,\,2,\,\ldots,$ and consider $\omega$ and $k$ as independent
quantities (see Refs.~\cite{21,22,23,24,25,26,27,28,29,30,31,32,34,35,36,37}
for details). When calculating the van der Waals (Casimir-Polder) forces
between separate atoms, the role of polarization tensor is played by the
atomic dynamic polarizabilities \cite{52}.

It is well known that the polarization tensor is closely related to the
in-plane and out-of-plane nonlocal conductivities of graphene
\cite{31,38,39,41,46}
\begin{eqnarray}
&&
\sila=-i\frac{\omega}{4\pi\hbar k^2}\Pi_{00}{\ar},
\nonumber \\
&&
\sita=i\frac{c^2}{4\pi\hbar\omega k^2}\Pi{\ar}.
\label{eq3}
\end{eqnarray}
\noindent
Substituting Eq.~(\ref{eq3}) in Eq.~(\ref{eq1}), we express the reflection
coefficients on a graphene sheet in terms of its conductivities
\begin{eqnarray}
&&
r_{\rm TM}(\omega,\theta_i)=
\frac{2\pi\cos\theta_i\sila}{c+2\pi\cos\theta_i\sila},
\nonumber \\
&&
r_{\rm TE}(\omega,\theta_i)=-
\frac{2\pi\sita}{c\cos\theta_i+2\pi\sita}.
\label{eq4}
\end{eqnarray}

The reflectances  of graphene for two independent polarizations of the
electromagnetic field are defined as
\begin{equation}
{\cal R}_{\rm TM(TE)}(\omega,\theta_i)=
|r_{\rm TM(TE)}(\omega,\theta_i)|^2.
\label{eq5}
\end{equation}
\noindent
Taking into account that both the polarization tensor and the conductivities
of graphene are the complex quantities, from Eqs.~(\ref{eq4}) and (\ref{eq5})
one obtains
\begin{eqnarray}
&&
{\cal R}_{\rm TM}(\omega,\theta_i)=
\frac{4\pi^2\cos^2\theta_i\left[{\rm Re}^2\sila+
{\rm Im}^2\sila\right]}{\left[c+2\pi\cos\theta_i{\rm Re}\sila\right]^2+
4\pi^2\cos^2\theta_i{\rm Im}^2\sila},
\nonumber \\
&&
{\cal R}_{\rm TE}(\omega,\theta_i)=
\frac{4\pi^2\left[{\rm Re}^2\sita+
{\rm Im}^2\sita\right]}{\left[c\cos\theta_i+2\pi{\rm Re}\sita\right]^2+
4\pi^2{\rm Im}^2\sita}
\label{eq6}
\end{eqnarray}

These equations take especially simple form in the case of normal incidence
$\theta_i=k=0$
\begin{equation}
{\cal R}(\omega)={\cal R}_{\rm TM}(\omega,0)={\cal R}_{\rm TE}(\omega,0)=
\frac{4\pi^2\left[{\rm Re}^2\sig+
{\rm Im}^2\sig\right]}{\left[c+2\pi{\rm Re}\sig\right]^2+
4\pi^2{\rm Im}^2\sig},
\label{eq7}
\end{equation}
\noindent
where $\sig\equiv\sigma_{\|}(\omega,0)=\sigma_{\bot}(\omega,0)$ is the
conductivity of graphene in the local limit.
In the framework of the Dirac model, this conductivity was found \cite{46}
using Eq.~(\ref{eq3}) and exact expressions for the polarization tensor of
graphene with arbitrary band gap $\Delta$ and chemical potential $\mu$ at
any temperature $T$. This opens the way for a sophisticated treatment of the
reflectance of graphene (\ref{eq6}) and (\ref{eq7})
over the entire application region of
the Dirac model, i.e., at all frequencies below 1--2~eV.

We begin with a brief summary of the exact results for $\sig$ at arbitrary
values of $T$, $\Delta$, and $\mu$ \cite{46}. The real part of the local
conductivity is given by
\begin{eqnarray}
&&
{\rm Re}\sig=\sigma_0
\theta(\ho-\Delta)\frac{(\ho)^2+\Delta^2}{2(\ho)^2}
\nonumber \\
&&~~~~
\times
\left(
{\rm tanh}\frac{\ho+2\mu}{4k_BT}+{\rm tanh}\frac{\ho-2\mu}{4k_BT}\right),
\label{eq8}
\end{eqnarray}
\noindent
where the universal conductivity $\sigma_0=e^2/(4\hbar)$,
$\theta(x)$ is the step function equal to unity for $x\geq 0$ and to zero
for $x<0$.
Note that Eq.~(\ref{eq8}) is interpreted as originating from the
interband transitions \cite{8}.
At moderate frequencies satisfying
the condition $\Delta\leq\ho\ll 2\mu$ the following asymptotic expression is valid:
\begin{equation}
{\rm Re}\sigma(\omega)=\sigma_0
\frac{(\ho)^2+\Delta^2}{(\ho)^2}
\frac{\ho}{4k_BT}{\rm sech}^2\frac{\mu}{2k_BT}.
\label{eq9}
\end{equation}

The exact expression for the imaginary part of the local conductivity of
graphene takes the form \cite{46}
\begin{equation}
{\rm Im}\sig=\frac{\sigma_0}{\pi}\left[
\frac{2\Delta}{\ho}-
\frac{(\ho)^2+\Delta^2}{(\ho)^2}
\ln\left|\frac{\ho+\Delta}{\ho-\Delta}\right|
+Y(\omega)\right],
\label{eq10}
\end{equation}
\noindent
where
\begin{eqnarray}
&&
Y(\omega)=2
\int_{\Delta/(\ho)}^{\infty}\!\!dt\sum_{\kappa=\pm 1}
\frac{1}{e^{\frac{\ho t+2\kappa\mu}{2k_BT}}+1}
\nonumber \\
&&~~~~~~
\times
\left[1+\frac{(\ho)^2+\Delta^2}{(\ho)^2}\,\frac{1}{t^2-1}\right].
\label{eq11}
\end{eqnarray}
\noindent
Under the conditions $k_BT\ll\mu$, $\Delta<2\mu$, and $\ho\ll2\mu$
Eqs.~(\ref{eq10}) and (\ref{eq11}) lead to the simple asymptotic result
\begin{equation}
{\rm Im}\sig=\frac{\sigma_0}{\pi}\left[
\frac{4\mu}{\ho}-4\ln2\,\frac{k_BT}{\ho}
-\frac{(\ho)^2+\Delta^2}{(\ho)^2}\ln\frac{2\mu+\ho}{2\mu-\ho}\right].
\label{eq12}
\end{equation}
\noindent
The first Drude-like term on the right-hand side of this equation was
interpreted as originating from the intraband transitions \cite{6}.

If, however, $k_BT\ll\mu$, but $\Delta>2\mu$ one finds
\begin{eqnarray}
&&
{\rm Im}\sig=\frac{\sigma_0}{\pi}\left[
\frac{2\Delta}{\ho} +16\frac{\Delta^2}{\Delta^2-(\ho)^2}
\frac{k_BT}{\ho}e^{-\frac{\Delta}{2k_BT}}{\rm cosh}\frac{\mu}{k_BT}
\right.
\nonumber \\
&&~~~~~~~~\left.
-\frac{(\ho)^2+\Delta^2}{(\ho)^2}
\ln\left|\frac{\ho+\Delta}{\ho-\Delta}\right|\right].
\label{eq13}
\end{eqnarray}
\noindent
This asymptotic expression is applicable at any frequency.
Note that Ref.~\cite{46}, where Eq.~(\ref{eq13}) was obtained, contains two
typos, i.e., indicates an incorrect factor $8\ln 2$ instead of 16 and
$k_BT$ instead of $2k_BT$ in the power of an exponent.

In the end of this section we note that a knowledge of the exact expressions
(\ref{eq8}) and (\ref{eq10}) for the local conductivity of graphene $\sig$
enables one to obtain the highly accurate approximate values for the TM and TE
reflectances at the arbitrary angle of incidence. To do so, we represent the
nonlocal in-plane and out-of-plane conductivities of graphene in the form \cite{46}
\begin{eqnarray}
&&
\sila=\sig+O\left(\frac{v_F^2}{c^2}\right),
\nonumber \\
&&
\sita=\sig+O\left(\frac{v_F^2}{c^2}\right),
\label{eq14}
\end{eqnarray}
\noindent
where the nonlocal corrections to local results are of the order of $10^{-5}$.
Substituting Eq.~(\ref{eq14}) in Eq.~(\ref{eq6}), we obtain
\begin{eqnarray}
&&
{\cal R}_{\rm TM}(\omega,\theta_i)=
\frac{4\pi^2\cos^2\theta_i\left[{\rm Re}^2\sig+
{\rm Im}^2\sig\right]}{\left[c+2\pi\cos\theta_i{\rm Re}\sig\right]^2+
4\pi^2\cos^2\theta_i{\rm Im}^2\sig}+
O\left(\frac{v_F^2}{c^2}\right),
\nonumber \\
&&
{\cal R}_{\rm TE}(\omega,\theta_i)=
\frac{4\pi^2\left[{\rm Re}^2\sig+
{\rm Im}^2\sig\right]}{\left[c\cos\theta_i+2\pi{\rm Re}\sig\right]^2+
4\pi^2{\rm Im}^2\sig}+
O\left(\frac{v_F^2}{c^2}\right).
\label{eq15}
\end{eqnarray}
\noindent
This result is applicable with exception of only extremely narrow vicinities of
frequencies $\omega_k$, where
${\rm Re}\sigma(\omega_k)={\rm Im}\sigma(\omega_k)=0$ holds (see Sec.~IV for a few
examples). At all other frequencies Eq.~(\ref{eq15}) provides an explicit
angular dependence of the reflectances of graphene already discussed previously in
Refs.~\cite{33,40}. Because of this, below we mostly concentrate on the impact
of chemical potential on the reflectance of graphene at the normal incidence.
Note also that in calculations of the van der Waals and Casimir forces one
should take into account the dependence of the conductivity of graphene on both
$\omega$ and $k$. This is because in the Lifshitz theory they are not
constrained by the mass shell equation.

\section{Impact of chemical potential in the case of gapless graphene}

We consider the reflectance of graphene with $\Delta=0$ at the normal incidence
given by Eq.~(\ref{eq7}). The real and imaginary parts of conductivity are
presented in Eqs.~(\ref{eq8}), (\ref{eq10}), and (\ref{eq11}) where one should
put $\Delta=0$. Using these equations, we first computed the reflectance of
graphene ${\cal R}$ at $T=300$~K as a function of frequency in the frequency
region from $2\times 10^{-5}$ to 1~eV belonging to the microwave and infrared
domains. The computational results are presented in Fig.~\ref{fg1} by the lines
1, 2, and 3 plotted for the chemical potential $\mu$ equal to 0.02, 0.1, and
0.2~eV, respectively.

As is seen in Fig.~\ref{fg1},
the reflectance of graphene reaches the minimum value at
the frequencies ${\ho}_m=0.068$, 0.16, and 0.32~eV for the lines 1, 2, and 3,
respectively. With increasing $\mu$, the minimum of ${\cal R}$ becomes more
pronounced. Note also that at all frequencies ${\rm Re}\sigma\neq 0$ and
strongly depends on both $\omega$ and $\mu$. In so doing we have
$2\mu<{\ho}_m<{\ho}_0$ where $\omega_0$ is the root of the imaginary part
of conductivity  ${\rm Im}\sigma(\omega_0)=0$. For the lines labeled 2 and 3
${\ho}_m<2\mu$ and $\omega_m\approx\omega_0$ hold. An impact of the chemical
potential on ${\cal R}$ is rather strong only within the frequency interval
from $2\times10^{-4}$ to 0.6~eV which mostly belongs to the infrared domain.
For $\ho<0.1$~eV the reflectance is larger for graphene with smaller chemical
potential.

Next, we consider the dependence of ${\cal R}$ on the frequency at lower
temperatures.
In Fig.~\ref{fg2}(a),  the lines 1, 2, and 3 show the
computational results at the liquid nitrogen temperature $T=77$~K
for the chemical potential $\mu$ equal to 0.02, 0.1, and 0.2~eV, respectively.
As is seen in Fig.~\ref{fg2}(a), the minimum values of the reflectance
are reached  at ${\ho}_m=0.034$, 0.16, and 0.33~eV for the lines 1, 2, and 3,
respectively. Thus, in comparison with Fig.~\ref{fg1} plotted at $T=300$~K,
the position of the first minimum is shifted to lower frequencies.
For all the three lines the inequality ${\ho}_m<2\mu$ holds and
$\omega_m\approx\omega_0$.

Similar results, but at the liquid helium temperature $T=4.4$~K, are
presented in Fig.~\ref{fg2}(b). Here, ${\ho}_m=0.034$, 0.165, and 0.335~eV
and, again,  ${\ho}_m<2\mu$  and
$\omega_m\approx\omega_0$ for all the three lines.
Note that for the lines 2 and 3 the minimum values of ${\cal R}$ equal to
$10^{-10}$ and $<10^{-17}$ cannot be shown in the used scale of
Fig.~\ref{fg2}(b). So small minimum  values of ${\cal R}$ are caused by the
fact that at very low temperatures under the condition $\ho<2\mu$ the
real part of conductivity becomes negligibly small (at $T=0$ and
$\ho<2\mu$ the exact equality ${\rm Re}\sigma=0$ is valid \cite{46}).
The distinctive feature of Fig.~\ref{fg2}(b), as compared to
Figs.~\ref{fg1} and \ref{fg2}(a), is the presence of small peaks at
$\ho=2\mu$ just after the points of each minimum of ${\cal R}$.
These peaks arise for the reason that at zero temperature
${\rm Im}\sigma\to\infty$ when $\ho\to 2\mu$ resulting in an unphysically
narrow maximum ${\cal R}=1$ \cite{46}.

For a gapless graphene under consideration in this section, the behavior
of its reflectance at sufficiently small $\omega$ can be found analytically.
For this purpose we consider first the imaginary part of conductivity.
{}From Eqs.~(\ref{eq10}) and (\ref{eq11})  one obtains
\begin{eqnarray}
&&
{\rm Im}\sig=\frac{2\sigma_0}{\pi}
\int_{0}^{\infty}\!\!dt\sum_{\kappa=\pm 1}
\frac{1}{e^{\frac{\ho t+2\kappa\mu}{2k_BT}}+1}
\nonumber \\
&&~~~~~~
\times
\left(1+\frac{1}{t^2-1}\right).
\label{eq16}
\end{eqnarray}

It is convenient to introduce the new integration variable
$u={\ho}t/(2k_BT)$ in the first contribution on the right-hand side of
Eq.~(\ref{eq16}), which is associated with unity in the brackets, and
leave the second contribution as is
\begin{eqnarray}
&&
{\rm Im}\sig=\frac{4k_BT\sigma_0}{\pi\ho}
\int_{0}^{\infty}\!\!du\left(
\frac{e^{-\frac{\mu}{k_BT}}}{e^u+e^{-\frac{\mu}{k_BT}}}+
\frac{e^{\frac{\mu}{k_BT}}}{e^u+e^{\frac{\mu}{k_BT}}}
\right)
\nonumber \\
&&~~~
+
\frac{2\sigma_0}{\pi}
\int_{0}^{\infty}\!\frac{dt}{t^2-1}\sum_{\kappa=\pm 1}
\frac{1}{e^{\frac{\ho t+2\kappa\mu}{2k_BT}}+1}.
\label{eq17}
\end{eqnarray}
\noindent
The first contribution on the right-hand side of Eq.~(\ref{eq17})
has the first-order pole at $\omega=0$ whereas the second contribution
converges
in the sense of the principal value. At  $\omega=0$ the second
contribution is proportional to
\begin{equation}
\dashint_{0}^{\infty}\frac{dt}{t^2-1}=0.
\label{eq18}
\end{equation}
\noindent
Because of this, one can neglect by the second contribution on
the right-hand side of Eq.~(\ref{eq17}), as compared to the
first one. Integrating the first contribution with respect to $u$,
we arrive at
\begin{eqnarray}
{\rm Im}\sig&\approx&\frac{4k_BT\sigma_0}{\pi\ho}
\left[\ln\left(1+e^{-\frac{\mu}{k_BT}}\right)+
\ln\left(1+e^{\frac{\mu}{k_BT}}\right)\right]
\nonumber \\
&=&\frac{8k_BT\sigma_0}{\pi\ho}
\ln\left(2\cosh\frac{\mu}{2k_BT}\right).
\label{eq19}
\end{eqnarray}

Now we take into account that in accordance with Eq.~(\ref{eq8})
${\rm Re}\sigma$ goes to zero with vanishing frequency (we recall that
$\Delta=0$ in this case). Then, for sufficiently small frequencies
satisfying the condition
\begin{equation}
\ho\ll8k_BT\ln\left(2\cosh\frac{\mu}{2k_BT}\right),
\label{eq20}
\end{equation}
\noindent
we obtain from Eqs.~(\ref{eq7}) and (\ref{eq19})
\begin{eqnarray}
{\cal R}&\approx&1-\left[\frac{c}{2\pi{\rm Im}\sig}\right]^2
\label{eq21} \\
&=&1-\left[
\frac{\ho}{4\alpha k_BT\ln\left(2\cosh\frac{\mu}{2k_BT}\right)}\right]^2.
\nonumber
\end{eqnarray}
\noindent
Here, $\alpha=e^2/(\hbar c)$ is the fine structure constant.

To find the application region of the asymptotic expression (\ref{eq21}),
we have compared the results of analytic calculations using this expression
with the computational results presented in Figs.~\ref{fg1} and \ref{fg2}.
Thus, for the lines 1, 2, and 3 presented in Fig.~\ref{fg1} ($T=300$~K)
the analytic results are in agreement with the numerical ones within 1\%
at all frequencies below 0.15, 0.5, and 1~meV, respectively.
The application region of Eq.~(\ref{eq21}) only slightly depends on the
value of temperature. Thus, for the lines 1, 2, and 3 in Fig.~\ref{fg2}(a)
($T=77$~K) the analytic results agree with the numerical ones at smaller
than 0.1, 0.5, and 0.9~meV  frequencies,  respectively.
At $T=4.4$~K the region of applicability of Eq.~(\ref{eq21})  is almost
unchanged.

We continue with the case of gapless graphene and compute the reflectance
given by Eqs.~(\ref{eq7}), (\ref{eq8}), (\ref{eq10}), and (\ref{eq11}) as
a function of chemical potential at different temperatures.
The computational results at $\ho=0.01$~eV are presented in Fig.~\ref{fg3}
by the lines 1, 2, and 3 at $T=300$, 77, and 4.4~K, respectively, for the
values of $\mu$ varying from 0 to 0.2~eV.
As is seen in this figure, on the condition  that $\mu<\ho/2=0.005$~eV
the reflectance of graphene is almost independent on the chemical potential.
For $\mu>0.07$~eV the computational results become nearly independent on the
temperature. For the lines 1 and 2 the smallest value of ${\cal R}$ is reached
at $\mu=0$. At the liquid helium temperature (the line 3) the minimum value
of ${\cal R}$
is reached at $\mu=0.006$~eV, i.e., under a condition $\ho<2\mu$
[see the discussion related to Fig.~\ref{fg2}(b)].

As a next step, we compute the reflectance of graphene as a function of the
chemical potential at higher frequency $\ho=0.2$~eV. The computational
results are presented in Fig.~\ref{fg4}(a)
by the lines 1, 2, and 3 at $T=300$, 77, and 4.4~K, respectively.
In Fig.~\ref{fg4}(b) a vicinity of the points of minimum reflectance is shown
on an enlarged scale for better visualization.
{}From Fig.~\ref{fg4}(a) it is seen that for $\mu<0.05$~eV the reflectance of
graphene is nearly independent on $\mu$.
The minimum values of the reflectance are reached at $\mu=0.13$, 0.128, and
0.12~eV for the lines 1, 2, and 3, respectively. In all these cases the
condition $\ho<2\mu$ holds. This explains why the minimum value becomes smaller
with decreasing temperature. At $T=4.4$~K the minimum value of ${\cal R}$ is equal
to $3\times 10^{-11}$. The nature of a peak in Figs.~\ref{fg4}(a) and \ref{fg4}(b)
arising on the line 3 ($T=4.4$~K) under the condition $2\mu=\ho$, i.e., at
$\mu=0.1$~eV, is the same as was discussed in relation to Fig.~\ref{fg2}(b).
It arises because at zero temperature the imaginary part of conductivity
turns to infinity when $2\mu=\ho$ resulting in ${\cal R}=1$.
Note also that for $\ho=0.2$~eV the reflectance becomes temperature-independent
at $\mu>0.18$~eV.

\section{Interplay between nonzero band gap and chemical potential}

In this section, we consider the reflectance of graphene in the most general case
of nonzero temperature, band gap and chemical potential. The numerical computations
are performed using Eqs.~(\ref{eq7}), (\ref{eq8}), (\ref{eq10}), and (\ref{eq11}).
The computational results for the reflectance as a function of frequency at
$T=300$~K, $\mu=0.02$~eV are presented in Fig.~\ref{fg5}(a) by the lines 1 and 2
for the band gap $\Delta=0.03$ and 0.05~eV, respectively. The vicinity of the points
of minimum and maximum reflectance is shown on an enlarged scale
in Fig.~\ref{fg5}(b). As can be seen in these figures, the presence of a nonzero
band gap leads to considerable changes as compared to the line 1 in Fig.~\ref{fg1}
computed at the same temperature and chemical potential, but with $\Delta=0$.
The most important novel feature is that under the condition $\ho=\Delta$ we have
${\rm Im}\sigma=\infty$ and, as a result, ${\cal R}=1$ [see Eqs.~(\ref{eq7})
and (\ref{eq10})]. The respective two peaks in Figs.~\ref{fg5}(a) and \ref{fg5}(b)
are unphysically narrow (this point is discussed in Ref.~\cite{41}).
Another distinctive feature of the case $\Delta\neq 0$ is that the minimum values
of ${\cal R}$ are reached exactly at the frequencies $\omega_0$ where
${\rm Im}\sigma(\omega_0)=0$. In Fig.~\ref{fg5} we have
$\ho_0=\ho_m=0.0287$~eV for the line 1 and
$\ho_0=\ho_m=0.0375$~eV for the line 2.
The respective minimum reflectance ${\cal R}(\omega_m)=0$ because
for both lines the inequality $\ho_m<\Delta$ is valid and, in accordance
with Eq.~(\ref{eq8}), it holds ${\rm Re}\sigma(\omega_m)=0$.
In the wide range of frequencies $\ho<0.01$~eV the reflectance of graphene
increases with decreasing band gap.

We now turn our attention to the discussion of the role of varying temperature in
the case of nonzero band gap. For this purpose, the lines 1 and 2 in
Fig.~\ref{fg6} show the computational results for the reflectance of graphene
with $\mu=0.02$~eV as a function of frequency
for the band gap $\Delta=0.03$ and 0.05~eV, respectively,
at (a) $T=77$~K and (b) $T=4.4$~K.
As is seen in Figs.~\ref{fg6}(a) and \ref{fg6}(b), under the condition
$\ho=\Delta$ the imaginary part of conductivity is infinitely large and
${\cal R}=1$. Similar to the case of $T=300$~K in Fig.~\ref{fg5}, this happens
at $\ho=0.03$~eV for the line 1 and
at $\ho=0.05$~eV for the line 2 in both  Figs.~\ref{fg6}(a) and \ref{fg6}(b).
The position of the points of minimum $\omega_m$, contrastingly, depends on
temperature.
For the lines 1 and 2 in Fig.~\ref{fg6}(a) ($T=77$~K), we have
$\ho_m=0.019$ and 0.014~eV, respectively
[we recall that here, again, the minimum value of reflectance is reached at
$\omega_m=\omega_0$ where ${\rm Im}\sigma(\omega_0)=0$].
Taking into account that $\ho_m<\Delta$, Eq.~(\ref{eq8}) leads to
${\rm Re}\sigma(\omega_m)=0$ and, thus, ${\cal R}(\omega_m)=0$.
In Fig.~\ref{fg6}(b) ($T=4.4$~K), $\ho_m=0.014$~eV for the line 1.
For the line 2 the minimum of reflectance is reached at a very small frequency.
In both cases ${\rm Re}\sigma(\omega_m)=0$ and ${\cal R}(\omega_m)=0$.
Comparing Figs.~\ref{fg5} and \ref{fg6}, one can conclude that with decreasing
temperature the position of the points of minimum shifts to smaller frequencies.
It should be noted also that there is a small peak on the line 1 in
Fig.~\ref{fg6}(b) at the frequency $\ho=2\mu=0.04$~eV.
It appears for the same reason as was explained above when discussing
Fig.~\ref{fg2}(b). In the case of gapped graphene this explanation is applicable
only under the condition $2\mu>\Delta$ satisfied for the line 1.
For the line 2 in Fig.~\ref{fg6}(b) it holds $2\mu<\Delta$ and no additional
peaks appear.

Some of the numerical results obtained above can be reproduced analytically using
the asymptotic expressions presented in Sec.~II. Thus, for the line 1 in
Fig.~\ref{fg6}(b) one has $k_BT=0.00038~\mbox{eV}\ll\mu=0.02$~eV and
$\Delta=0.03~\mbox{eV}<2\mu=0.04$~eV. Because of this, the asymptotic
expression (\ref{eq12}) should be applicable in this case at sufficiently low
frequencies. In fact Eq.~({\ref{eq7}) for ${\cal R}$ with ${\rm Re}\sigma=0$
and ${\rm Im}\sigma$ given by Eq.~({\ref{eq12}) agrees with the results of
numerical computations to within 1\% at $\ho<0.01$~meV.

As one more example, we consider the line 2 in Fig.~\ref{fg6}(b).
Here, again, $k_BT\ll\mu=0.02$~eV, but $\Delta=0.05~\mbox{eV}>2\mu$.
Because of this, the asymptotic expression (\ref{eq13}) is applicable.
One can check that the substitution of Eq.~(\ref{eq13}) in Eq.~({\ref{eq7})
leads to the values of ${\cal R}$ in agreement to within 1\% with the results
of numerical computations in the wide frequency regions
$\ho<5\times10^{-11}$~eV and $0.2~\mbox{meV}<\ho<0.05$~eV (we recall that for these
frequencies ${\rm Re}\sigma$ is again equal to zero).

The asymptotic expression (\ref{eq13}) also allows to estimate the value
of frequency where the reflectance of graphene vanishes:
${\cal R}(\omega_m)=0$. In this case
$\omega_m=\omega_0$ where ${\rm Im}\sigma(\omega_0)=0$.
Taking into account that for the line 2
we have $\ho_0\ll\Delta$ [in Fig.~\ref{fg6}(b)
the region of so low frequencies is not shown], one can expand in
Eq.~(\ref{eq13}) up to the first power in the small parameter $\ho/\Delta$
and obtain
\begin{equation}
{\rm Im}\sig\approx\frac{\sigma_0}{\pi}\left(
-\frac{8}{3}\,\frac{\ho}{\Delta} +
\frac{16k_BT}{\ho}e^{-\frac{\Delta}{2k_BT}}{\rm cosh}\frac{\mu}{k_BT}
\right).
\label{eq22}
\end{equation}
\noindent
The root of this expression is equal to
\begin{equation}
\ho_0=\left(6k_BT\Delta
e^{-\frac{\Delta}{2k_BT}}{\rm cosh}\frac{\mu}{k_BT}
\right)^{1/2}.
\label{eq23}
\end{equation}
\noindent
For the values all the parameters under consideration this results in
$\ho_0\approx 0.01$~meV. Although in a vicinity of the point of minimum
${\cal R}$ the asymptotic expression is not as exact, as in the frequency
regions indicated above, the obtained value is in qualitative agreement
with $\ho_0= 0.015$~meV found by means of numerical computations.

In the end of this section we consider the dependence of ${\cal R}$ on the
chemical potential for the gapped graphene at different temperatures.
In Fig.~\ref{fg7} the computational results are presented for the reflectance
of graphene with $\Delta=0.1$~eV at the frequency $\ho=0.01$~eV as a function
of $\mu$. The lines 1, 2, and 3 are plotted at $T=300$, 77, and 4.4~K,
respectively. As is seen in Fig.~\ref{fg7}, the lines 2 and 3
reach their minimum values at $\mu=0.035$ and 0.0506~eV, respectively.
For these values of $\mu$ it holds ${\rm Im}\sigma(0.01~\mbox{eV}/\hbar)=0$
at $T=77$ and 4.4~K, respectively. Taking into account also that
$\ho=0.01~\mbox{eV}<\Delta$, one concludes that  ${\rm Re}\sigma=0$ and, thus,
${\cal R}=0$. From the comparison of Figs.~\ref{fg7} and \ref{fg3}, we find
that the presence of a nonzero band gap results in a considerable decrease
of graphene reflectance.

The obtained results strongly depend on the specific values of frequency and
band gap. In Fig.~\ref{fg8}(a) the reflectance of gapped graphene
with $\Delta=0.1$~eV (the same as in Fig.~\ref{fg7}), but at the
frequency $\ho=0.07$~eV, is plotted as a function
of $\mu$ by the lines 1, 2, and 3 at $T=300$, 77, and 4.4~K,
respectively.
In this case all the three lines are of the same character, and the lines 2 and 3
are almost coinciding. In the region $2\mu<\Delta=0.1$~eV at $T=77$ and 4.4~K
the reflectance is nearly independent on $\mu$. The minimum values of ${\cal R}$
on the lines 1, 2, and 3 are reached at $\mu=0.0726$, 0.0791, and 0.0787,
respectively. For these values of $\mu$ we have
${\rm Im}\sigma(0.07~\mbox{eV}/\hbar)=0$
at $T=300$, 77, and 4.4~K, respectively. The minimum reflectance vanishes,
${\cal R}=0$, because $\ho=0.07~\mbox{eV}<\Delta$ and, thus,
${\rm Re}\sig=0$.

Note that in Figs.~\ref{fg5}--\ref{fg7} and \ref{fg8}(a) plotted for a gapped
graphene the minimum  reflectance is ${\cal R}=0$. This means that in the
narrow vicinities of points, where the minimum is reached, one cannot use
Eq.~(\ref{eq15}) for calculation of the angular dependence of the TM and TE
reflectances. For this purpose the nonlocal corrections to the conductivity of
graphene would be required. Equation (\ref{eq15}), however, still stands  over the
entire region of frequencies with exception of only
narrow vicinities of the points
of minimum reflectance.

Finally, in Fig.~\ref{fg8}(b) we plot the reflectance of gapped graphene
with $\Delta=0.04$~eV at $\ho=0.07$~eV as a function
of $\mu$. The three lines 1, 2, and 3 are again plotted at $T=300$, 77, and 4.4~K,
respectively. All the three lines in  Fig.~\ref{fg8}(b) differ qualitatively.
Unlike the previous figure, now $\ho>\Delta$ which means that ${\rm Re}\sigma\neq 0$
[see Eq.~(\ref{eq8})]. As a result, the minimum values of ${\cal R}$ are not equal to
zero, but are determined by the value of ${\rm Re}\sigma(0.07~\mbox{eV}/\hbar)$.
[We recall that at the point of minimum ${\rm Im}\sigma(0.07~\mbox{eV}/\hbar)=0$.]
At  $T=300$, 77, and 4.4~K, i.e., for the lines 1, 2, and 3, the respective minimum
values of  ${\cal R}$ are equal to $4.8\times 10^{-5}$, $4.5\times 10^{-6}$, and
$1.2\times 10^{-11}$. They are reached at $\mu=0.044$, 0.048, and 0.04455~eV,
respectively. A small peak on the line 3 ($T=4.4$~K) arises at
$\mu=\ho/2=0.035$~eV for the reasons discussed above.

\section{Conclusions and discussion}

   In the foregoing, we have investigated the impact of
chemical potential on the reflectance of graphene in the
application region of the Dirac model, i.e., in the
microwave and infrared domains. For this purpose, the
reflectance was expressed via the polarization tensor or,
alternatively, via the conductivity of graphene. Both these
quantities have been found recently for arbitrary values of
the chemical potentials and band-gap parameter at any
temperature in the framework of the Dirac model basing on
the first principles of quantum electrodynamics.

The obtained expressions were used to perform numerical
computations of the graphene reflectance at the normal
incidence as a function of frequency and chemical potential
at different temperatures. Both the cases of gapless and
gapped graphene were considered. It is shown that the
major impact of chemical potential on the graphene
reflectance occurs in the domain of infrared optics. In
this domain the reflectance possesses the minimum values
reached at the roots of the imaginary part of the
conductivity of graphene. The magnitudes of the minimum
values of reflectance are either equal to zero or
determined by the real part of conductivity if the latter
is nonzero. The reflectance of gapped graphene is lower
than that of a gapless one. Over the wide range of
frequencies, the reflectance of graphene increases with
increasing chemical potential and decreasing band gap.
We have also demonstrated that the reflectance of gapped
graphene possesses narrow peaks and turns to unity at
the frequencies where the imaginary part of conductivity
turns to infinity. By and large the reflectance of
graphene is determined by the values of frequency,
temperature, band-gap parameter, chemical potential and
their interplay. The results of numerical computations
and analytic asymptotic calculations were compared and
found to be in good agreement.
At higher frequencies and, specifically, for visible light,
some theoretical approaches beyond the scope of the Dirac
model are developed (see Sec.~I and, e.g., Refs.~\cite{53,54,55}
where small quadratic in momentum subdominant contributions
to a dominant linear Dirac Hamiltonian are considered).

To conclude, the developed complete theory of the
reflectance of graphene takes into account the combined
effect of its realistic properties, such as nonzero
chemical potential and band-gap parameter in the application
region of the Dirac model.
The obtained results could
be interesting not only for fundamental physics, but for
the listed above numerous technological applications
of graphene.


\section*{Acknowledgments}
The work of V.M.M. was partially supported by the Russian
Government
Program of Competitive Growth of Kazan Federal University.

\newpage
\begin{figure}[b]
\vspace*{-10cm}
\centerline{\hspace*{2.5cm}
\includegraphics{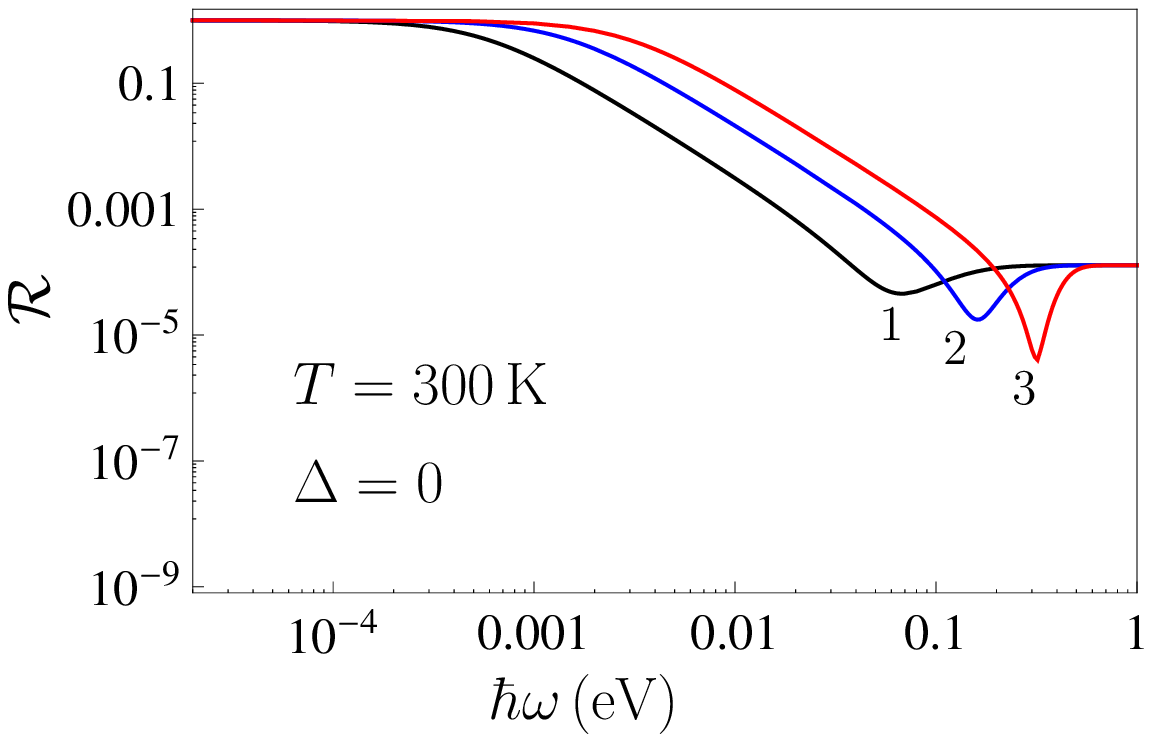}
}
\vspace*{-9cm}
\caption{\label{fg1}
The reflectance of gapless graphene at $T = 300$~K is
shown as a function of frequency by the lines 1, 2, and
3 for the chemical potential $\mu = 0.02$, 0.1, and
0.2~eV, respectively.
}
\end{figure}
\begin{figure}[b]
\vspace*{-0cm}
\centerline{\hspace*{2.5cm}
\includegraphics{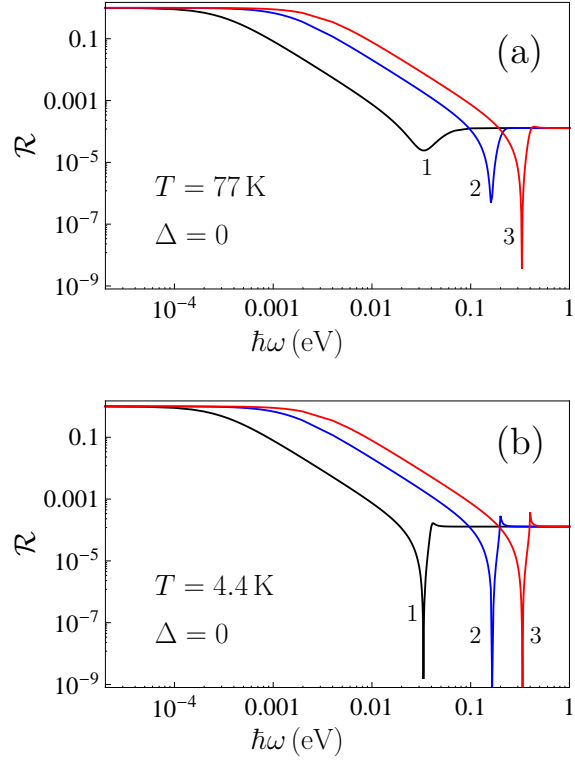}
}
\vspace*{-16cm}
\caption{\label{fg2}
The reflectance of gapless graphene at (a) $T = 77$~K
and (b) $T = 4.4$~K is shown as a function of frequency
by the lines 1, 2, and 3 for the chemical potential
$\mu = 0.02$, 0.1, and 0.2~eV, respectively.
}
\end{figure}
\begin{figure}[b]
\vspace*{-10cm}
\centerline{\hspace*{2.5cm}
\includegraphics{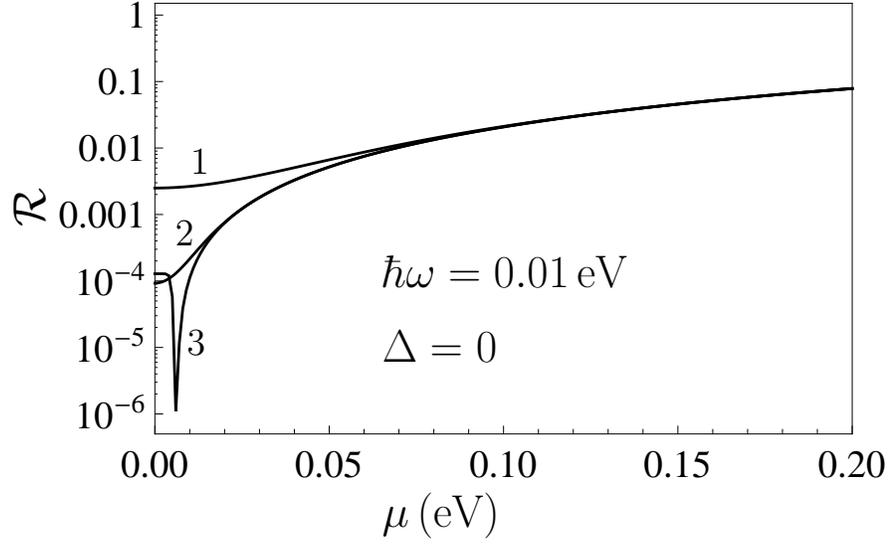}
}
\vspace*{-9cm}
\caption{\label{fg3}
The reflectance of gapless graphene at
$\hbar \omega =0.01$~eV
is shown as a function of the chemical potential
by the lines 1, 2, and 3 at $T = 300$, 77, and 4.4~K,
respectively.
}
\end{figure}
\begin{figure}[b]
\vspace*{-0cm}
\centerline{\hspace*{2.5cm}
\includegraphics{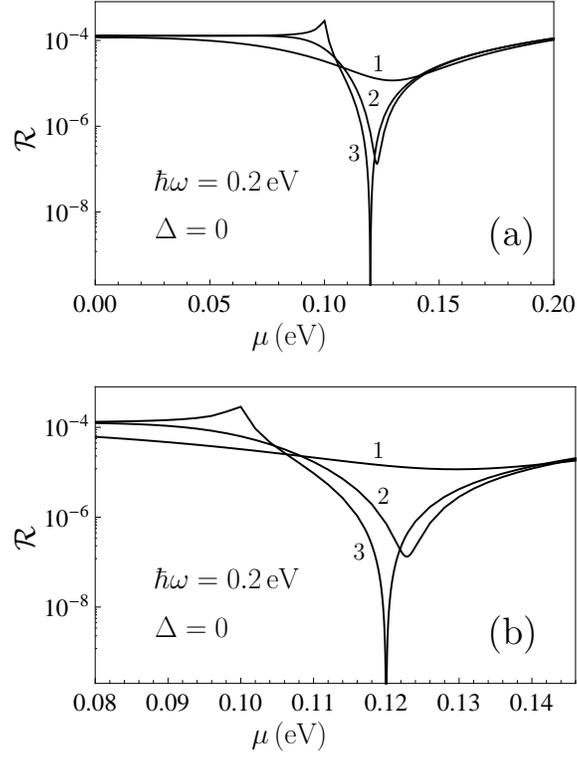}
}
\vspace*{-16cm}
\caption{\label{fg4}
(a) The reflectance of gapless graphene at
$\hbar \omega =0.2$~eV
is shown as a function of the chemical potential
by the lines 1, 2, and 3 at $T = 300$, 77, and 4.4~K,
respectively.
(b) A vicinity of the points of minimum
reflectance is shown on an enlarged scale.
}
\end{figure}
\begin{figure}[b]
\vspace*{-0cm}
\centerline{\hspace*{2.5cm}
\includegraphics{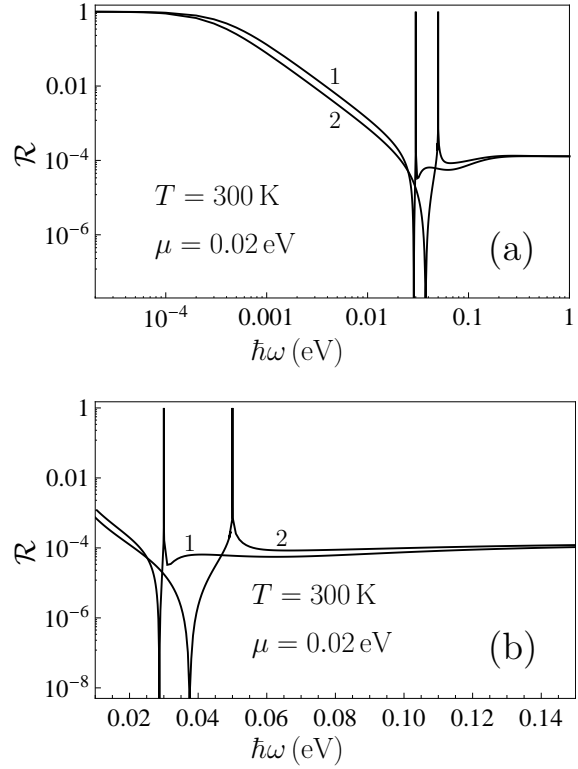}
}
\vspace*{-16cm}
\caption{\label{fg5}
(a) The reflectance of gapped graphene with $\mu = 0.02$~eV
at $T = 300$~K is shown as a function of frequency
by the lines 1 and 2 for the band gap $\Delta = 0.03$ and
0.05~eV, respectively. (b) A vicinity of the points of
minimum reflectance is shown on an enlarged scale.
}
\end{figure}
\begin{figure}[b]
\vspace*{-0cm}
\centerline{\hspace*{2.5cm}
\includegraphics{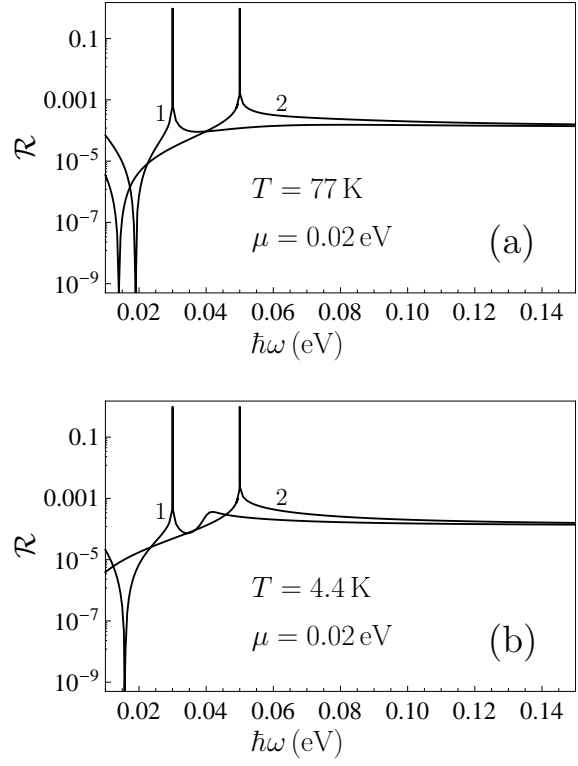}
}
\vspace*{-16cm}
\caption{\label{fg6}
The reflectance of gapped graphene with $\mu = 0.02$~eV
at (a) $T = 77$~K and (b) $T = 4.4$~K
is shown as a function of frequency
by the lines 1 and 2 for the band gap $\Delta = 0.03$ and
0.05~eV, respectively.
}
\end{figure}
\begin{figure}[b]
\vspace*{-10cm}
\centerline{\hspace*{2.5cm}
\includegraphics{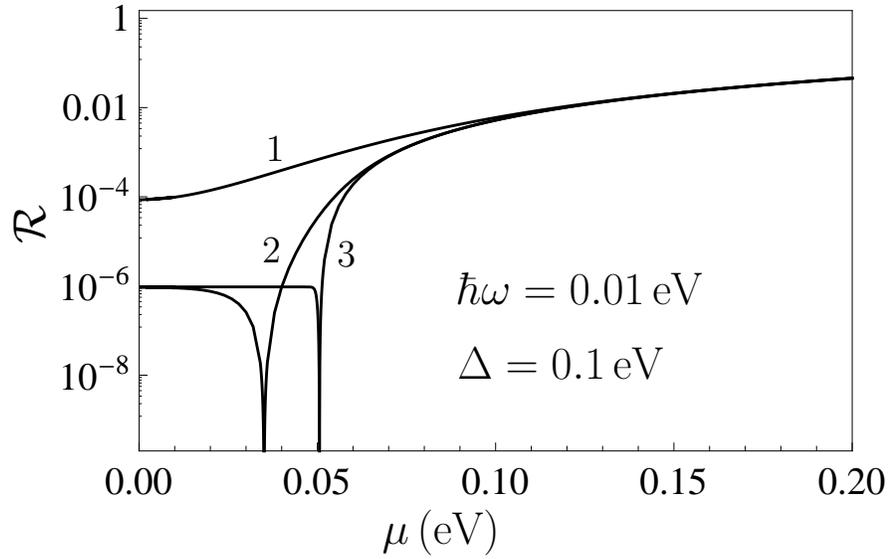}
}
\vspace*{-9cm}
\caption{\label{fg7}
The reflectance of gapped graphene with $\Delta = 0.1$~eV at
$\hbar \omega = 0.01$~eV is shown as a function of the chemical
potential by the lines 1, 2, and 3 at $T = 300$, 77, and 4.4~K,
respectively.
}
\end{figure}
\begin{figure}[b]
\vspace*{-0cm}
\centerline{\hspace*{2.5cm}
\includegraphics{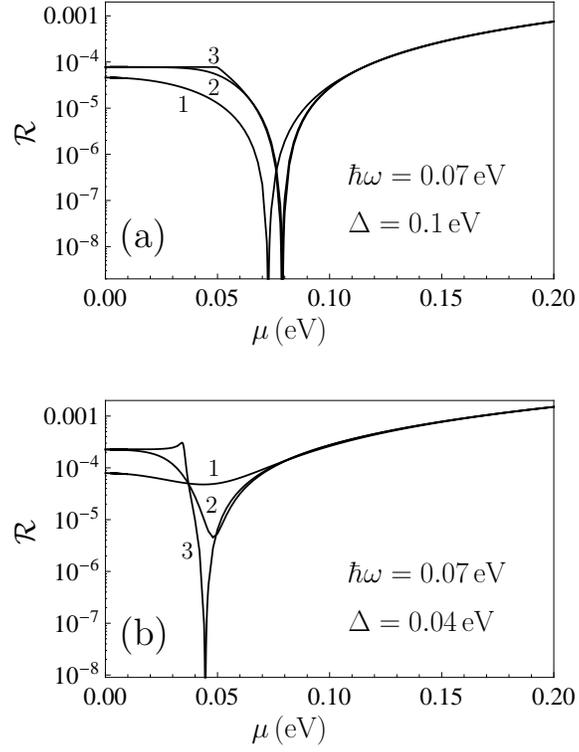}
}
\vspace*{-16cm}
\caption{\label{fg8}
The reflectance of gapped graphene with (a) $\Delta = 0.1$~eV
and (b) $\Delta = 0.04$~eV at
$\hbar \omega = 0.07$~eV is shown as a function of the chemical
potential by the lines 1, 2, and 3 at $T = 300$, 77, and 4.4~K,
respectively.
}
\end{figure}
\end{document}